\documentclass[journal, 10pt, twoside]{IEEEtran}
\IEEEoverridecommandlockouts

\usepackage{cite}
\usepackage{amsmath,amssymb,amsfonts}
\usepackage{algorithm}
\usepackage{algorithmicx}
\usepackage[]{algpseudocodex}
\usepackage{graphicx}
\usepackage{tabularx}
\usepackage{textcomp}
\usepackage{xcolor}
\usepackage{subcaption}
\def\BibTeX{{\rm B\kern-.05em{\sc i\kern-.025em b}\kern-.08em
    T\kern-.1667em\lower.7ex\hbox{E}\kern-.125emX}}

\usepackage[nolist,nohyperlinks]{acronym}

\usepackage{booktabs}
\usepackage{svg}
\usepackage{cleveref}
\usepackage{url}

\newcommand{\papertitle}{A Microservice-Based Platform for Sustainable and Intelligent SLO Fulfilment and Service Management}
\begin{document}

\title{\papertitle}

\author{Juan~Luis~Herrera, Daniel~Wang, and Schahram~Dustdar\vspace{-1cm}\thanks{
This work has been submitted to the IEEE for possible publication. Copyright may be transferred without notice, after which this version may no longer be accessible. This work has been funded by the European Union under the MSCA project RENOS (contract number 101205037). Views and opinions expressed are however those of the authors only and do not necessarily reflect those of the European Union or the European Research Executive Agency. Neither the European Union nor the granting authority can be held responsible for them. \textbf{(Corresponding author: Juan Luis Herrera.)}}%
\thanks{Juan Luis Herrera, Daniel Wang, and Schahram Dustdar are with the Distributed Systems Group, TU Wien, Austria (e-mail: j.gonzalez@dsg.tuwien.ac.at). Schahram Dustdar is also with ICREA in Spain.}%
\thanks{Digital Object Identifier 00.000/TSC.0000.00000000}
}

\markboth{IEEE Transactions on Services Computing, Vol. 0, No. 0, January 0000}{J.L. Herrera \MakeLowercase{\textit{et al.}}: \papertitle}


\maketitle

\begin{abstract}
The Microservices Architecture (MSA) design pattern has become a staple for modern applications, allowing functionalities to be divided across fine-grained microservices, fostering reusability, distribution, and interoperability. As MSA-based applications are deployed to the 
Computing Continuum (CC), meeting their 
Service Level Objectives (SLOs) 
becomes a challenge. Trading off performance and sustainability SLOs 
is especially challenging. This challenge can be addressed with intelligent decision systems, able to reconfigure the services during runtime to meet the SLOs. However, developing these agents while adhering to the MSA pattern is complex, especially because CC providers, who have key know-how and information to fulfill these SLOs, must comply with the privacy requirements of application developers. This work presents the Carbon-Aware SLO and Control plAtform (CASCA), an open-source MSA-based platform that allows CC providers to reconfigure services and fulfill their SLOs while maintaining the privacy of developers. CASCA is architected to be highly reusable, distributable, and easy to use, extend, and modify. 
CASCA has been evaluated in a real CC testbed for a media streaming service, where decision systems implemented in Bash, Rust, and Python successfully reconfigured the service, unaffected by upholding privacy. 
\end{abstract}


\section{Introduction}\label{sec:intro}


The Service-Oriented Architecture (SOA) has revolutionized the paradigm for software development and operation, 
dividing applications into multiple artifacts or \textit{services}, each containing a subset of the application's functionality~\cite{DBLP:journals/vldb/PapazoglouH07}. 
Services are loosely coupled, can be deployed to multiple, distributed machines, and can be \textit{composed} to perform complex functionalities. 
Moreover, SOAs allow services to be reused across applications, and decouple the 
technologies used for their implementations. 
These benefits are increased by the Microservices Architecture (MSA) design pattern, which proposes that services in SOAs should become fine-grained \textit{microservices}~\cite{DBLP:journals/jss/Waseem0S20}. 
MSAs allow SOAs to be further distributed and modularized, emphasizing their benefits.

Within SOAs and MSAs, the performance of each service is important, as it may affect the performance of other services during composition. In 
the Computing Continuum (CC) paradigm, the desired performance of a service is declared as its Service Level Objectives (SLOs)~\cite{DBLP:conf/applied/Casamayor-Pujol24,DBLP:conf/sysose/SedlakCDD24}. 
For example, an application operator may declare SLOs for maximum response time, minimum availability or thresholds on energy consumption~\cite{DBLP:conf/applied/Casamayor-Pujol24}. 
Given the increasing relevance of sustainability, including government policies such as Sustainable Development Goals~\cite{SDGs} or the European Green Deal~\cite{EuropeanGreenDeal}, and industry and academia movements like green computing~\cite{DBLP:conf/padl/FortiB22,DBLP:journals/access/PaulSABBRAAM23}, makes sustainability a key SLO. Unlike energy consumption, sustainability has a greater focus on carbon-awareness,  
considering the carbon intensity of the energy consumed~\cite{DBLP:conf/padl/FortiB22,DBLP:journals/tps/RadovanovicKSCDR22}.

An important aspect of fulfilling these SLOs is the configuration of each service~\cite{DBLP:journals/evs/SedlakCMDD25}. 
For example, the number of threads or worker processes used, the level of quality, or hardware acceleration may be configured to reduce the carbon footprint of the service or to increase its performance. Nonetheless, configuring services to achieve specific SLOs is complex~\cite{DBLP:conf/sysose/SedlakCDD24,DBLP:journals/evs/SedlakCMDD25}. On the one hand, the effect of a given service configuration on its SLOs is complicated to determine, and thus, so is deriving an appropriate configuration to meet the specified SLOs, especially in dynamic conditions. 
On the other hand, some SLOs can be non-orthogonal or even in conflict, further complicating their fulfilment, e.g., activating GPU acceleration can increase the throughput of the application, but it will also increase its energy consumption. 
State-of-the-art works propose leveraging multiple techniques to handle and adapt this configuration over time, including dynamic optimization~\cite{DBLP:journals/tsc/LuWLWLF24}, heuristic algorithms~\cite{DBLP:conf/cnsm/FurstA0P18}, or artificial intelligence (AI)-based techniques~\cite{DBLP:conf/sysose/SedlakCDD24, DBLP:journals/evs/SedlakCMDD25}, such as Deep Reinforcement Learning (DRL). These techniques are better suited for different use cases, as traditional optimization and heuristics are lightweight, but require parameter tuning, 
while AI-based systems, after the resource-intensive a training process, can learn autonomously. 
As a result, different decision systems to adapt the configuration to meet specific SLOs can be developed based on these various techniques. 

However, the development and usage of a decision system for these tasks is complex due to the need for interoperability. 
Information about SLO fulfilment may come from multiple sources, such as the service's database, metrics measured by clients, 
hardware (e.g., energy consumption meters), or external sources (e.g., governmental data about the energy mix used), which are not always mutually observable by an external system. 
Furthermore, the ideal developers for these decision systems are the infrastructure providers, such as CC providers, given their know-how in service management and the amount of SLO-related data they have. However, infrastructure providers must adhere to privacy requirements that avoid them from having observability over the services running in their infrastructure, allowing, e.g., Netflix to be hosted in AWS, despite Amazon Prime Video being their competitor~\cite{NetflixAWS}. Hence, it is also necessary to implement measures to ensure infrastructure providers can develop and use these decision systems without providing them knowledge about the service itself. Finally, implementing solutions for these difficulties at the decision systems themselves makes them become coarse-grained and address many concerns at once, in direct opposition to the principles of SOAs and MSAs.

To address these difficulties, we propose the Carbon-Aware SLO and Control plAtform (CASCA) to 
enable the development and use of decision systems, based on both traditional and AI techniques, for SLO fulfilment and service management. CASCA has been designed according to the principles of MSAs, ensuring each architectural module can be implemented as a loosely-coupled and fine-grained service, and facilitating distribution and reuse. Moreover, CASCA has been designed to be easy to extend and adapt, and offers multiple declarative interfaces to maximize its flexibility and accelerate the modification, removal, or addition of data sources or SLOs. CASCA ensures privacy by allowing service operators to remove semantics on SLO declarations and service control. 
Furthermore, CASCA includes the Energy Mix Manager (EMMA), a microservice to easily integrate carbon intensity into the MSA's service management. The main contributions of this work are:
\begin{itemize}
    \item The proposal of the CASCA platform for SLO fulfilment and service management. CASCA 
    fosters reuse, adaptation, extension, and distribution.
    \item The declaration of APIs for service control and SLO fulfilment with CASCA. These APIs adhere to open standards and are privacy-protecting.
    \item The proposal of the EMMA microservice, providing a unified interface for gathering carbon intensity data.
    \item The implementation of CASCA in a media streaming service use case. 
    Every component used in this use case, including CASCA, is open source to foster replication.
    \item The evaluation of CASCA in a real CC testbed, leveraging real implementations for the platform, thus 
    showing the applicability of CASCA to real scenarios.
\end{itemize}

The remainder of this manuscript is structured as follows. \Cref{sec:related} presents background and related works. 
\Cref{sec:architecture} presents the architecture of CASCA, which is then applied to the media streaming use case in \Cref{sec:use-case-implementation}. This implementation is evaluated in \Cref{sec:evaluation}. Finally, \Cref{sec:conclusion} concludes the work and presents future research lines.
\section{Related works}\label{sec:related}

The problem of continuously adapting the configuration of a service is of key importance in the CC, and thus, research works have studied diverse methods to perform it. 
Fürst \textit{et al.} present a heuristic for adaptative service configuration, named \textit{Simplified Code Selection}~\cite{DBLP:conf/cnsm/FurstA0P18}. This heuristic observes the behaviour of a service over a selectable time window, and contrasts it with performance metrics, degrading the quality if more performance is required and upgrading the quality if stable and good performance is achieved. 
This approach, while interesting, requires profound knowledge on how specific values for the parameters of the application affect performance. Shirkoohi \textit{et al.} presented another decision system, aimed to the domain of video inference~\cite{DBLP:conf/iccv/ShirkoohiHNA21}. Their work proposes Adaptive Model Streaming, in which a knowledge distillation process for 
computer vision models is configured to optimize 
model size and bandwidth usage. 
Automated configuration is achieved through the implementation of a gradient guided-based system, which selects the most relevant parameters over time. 
Adaptive Model Streaming does not require 
a priori knowledge of configuration-SLO interactions, but requires deep knowledge about the computer vision models instead.

As means to overcome this need for knowledge, AI-based decision systems have been proposed. As a representative of these approaches, Lapkovskis \textit{et al.} proposed using a variety of AI-based methods to perform reconfiguration, including Active Inference and DRL techniques\cite{DBLP:conf/edge/LapkovskisSMDD25}. These techniques were used to configure a video conferencing application, 
aimed at fulfilling multiple SLOs, including latency and throughput, as well as SLOs related to minimum video quality, with no prior knowledge. 
However, the proposed methodology requires an ad-hoc implementation of the service that 
decision systems can configure and observe. 
RL-Adapt~\cite{DBLP:journals/cn/CaoDSZD23} also leverages DRL 
to adjust the configuration of a video streaming service. RL-Adapt considers the bandwidth fluctuations of the system and can handle multiple types of video streaming configuration parameters, 
as well as custom reward functions. 
The presented proof of concept is applied to pedestrian detection, autopilot, and stream popularity. However, RL-Adapt is aimed at stream processing specifically, as well as explicitly for DRL-based decision systems, limiting the capabilities of developing other decision systems.

A key aspect throughout these works are SLOs: all these decision systems focus on fulfilling the SLOs by reconfiguring these services. 
Casamayor \textit{et al.} proposed a framework, named DeepSLOs, for SLO categorization~\cite{DBLP:conf/applied/Casamayor-Pujol24}. They propose that stakeholders do not directly declare 
\textit{low-level SLOs}, but 
\textit{high-level SLOs} that express the desired status of the service. 
At the same time, these SLOs involve semantics from different aspects of the system, which they separate into 
\textit{infrastructural} SLOs, 
that refer to metrics observable by the infrastructure provider regardless of the service being executed, and 
\textit{service} SLOs, 
semantically important and observable by the service user and developer, but not from the infrastructure provider. This framework was ultimately implemented by Sedlak \textit{et al.} through SLO diffusion~\cite{DBLP:conf/sysose/SedlakCDD24}. Their main result is the use of casual probabilistic models, namely Bayesian Networks, to transform high-level SLOs into low-level SLOs. 
These works set the direction for CASCA's SLO considerations: while service developers focus on service SLOs, CC providers set the infrastructural SLOs, and they are not observable for one another. Hence, given the role of CC providers, they should be given observability, but not context, of service SLOs. 

Given the interest in sustainability, there has also been a transition from the traditional energy consumption SLOs towards carbon footprint SLOs. This is reflected by the efforts on carbon-aware computing, such as Google's~\cite{DBLP:journals/tps/RadovanovicKSCDR22}. This work, authored by Radovanović \textit{et al.}, focuses on how energy-awareness and carbon-awareness are different concerns, as carbon-awareness requires conscience of the carbon footprint as well as of energy consumption. Hence, 
it is possible to reduce the overall carbon footprint of the infrastructure, even if the total energy consumption does not vary. 
Carbon-awareness thus is ultimately a high-level infrastructural SLO, but unlike usual infrastructural SLOs, it depends on 
the carbon intensity of energy. Supporting carbon-awareness as an SLO within a platform thus calls for streamlining the observability of the carbon intensity of the infrastructure within the platform. These considerations are a foundation CASCA is built upon. 

As part of these efforts, frameworks have been developed to enable service reconfiguration and SLO fulfilment, being the most similar works to CASCA. Fürst \textit{et al.} present DivProg~\cite{DBLP:conf/cnsm/FurstA0P18}, in the same work as Simplified Code Selection, as an approach to allow external systems to configure the services executing in the CC. Service developers can include DivProg annotations in their functions to specify different configurations and their effects on SLOs. 
On the one hand, requiring source code modifications restricts the use of off-the-shelf services. On the other hand, only service SLOs can be expressed with DivProg, not infrastructural SLOs. 
Cao \textit{et al.} also include a programming model for their already discussed RL-Adapt approach~\cite{DBLP:journals/cn/CaoDSZD23}. RL-Adapt's programming model is based on Python, and allows a user to declaratively reconfigure the service, as well as calculate the reward of the target DRL agent. 
RL-Adapt's declarative approach is based on specifying the possible configurations that the agent may control and how to evaluate the result of a configuration. 
RL-Adapt's model 
also focuses on service SLOs first and foremost. 
Moreover, it is tightly coupled with the DRL and stream processing libraries of RL-Adapt. These frameworks show the need to build a platform for service reconfiguration and SLO fulfilment, but are tightly coupled to the methods presented in the same works. 
\section{CASCA architecture}\label{sec:architecture}



This section presents the main elements of CASCA: \Cref{subsec:telemetry} details the SLO-related elements, including EMMA, while \Cref{subsec:service-api} focuses on the service API and its controlling aspects, 
and \Cref{subsec:ai} describes the integration with decision systems.

\begin{figure*}
    \centering
    \includegraphics[width=0.7\textwidth]{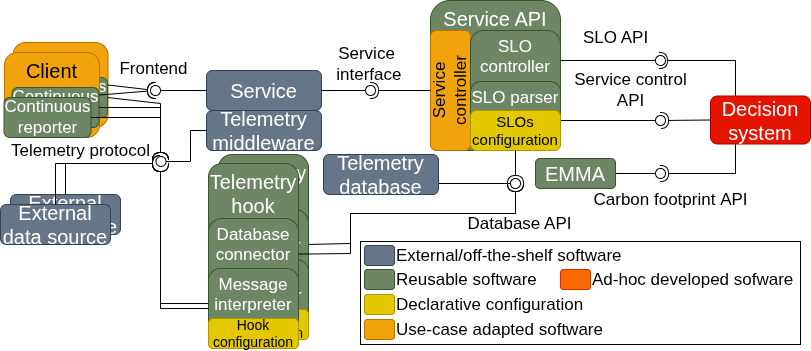}
    \caption{Complete architecture of the CASCA platform}
    \label{fig:architecture}
\end{figure*}

\subsection{SLOs and telemetry}\label{subsec:telemetry}

One of CASCA's objectives is to simplify SLO fulfilment in SOAs and MSAs. This simplification can be structured into three main aspects: gathering SLO-related information from multiple sources, 
unifying the information, and simplifying the configuration of SLOs over this information. 
While the latter 
is part of the service API, the two former aspects are addressed by the SLO and telemetry modules. These modules include adapted clients, external data sources, telemetry middlewares, the telemetry database, and telemetry hooks. 

The workflow of data through CASCA begins by setting up, at least, one \textit{telemetry middleware}. This middleware must be able to receive network messages containing the telemetry data from any number of sources, and forward these messages to any number of services. 
A message-oriented middleware broker that supports standard or open protocols (e.g. MQTT) is ideal to serve as a middleware, although other middlewares can be used. 
Moreover, this telemetry middleware does not need to be especially adapted or tailored to CASCA
An important aspect of the telemetry middleware is that, while \Cref{fig:architecture} only shows one telemetry middleware instance for simplicity, CASCA supports the instantiation of multiple, different telemetry middlewares, if so necessary. 

\Cref{fig:architecture} shows \textit{external data sources}, which represent 
elements that are not directly part of the application, but report data that can be considered relevant to SLOs, such as 
hardware devices, service orchestrators, or virtualization platforms. 
There is no assumption or constraint to the architecture of external data sources. 
An exception to this rule is the \textit{client}, 
the software leveraged by users to access a frontend. Clients include mobile phone applications, web browsers, or desktop applications. 
The client service is important, 
as it is the only service able to report the quality experienced at the client-side. 
If the existing clients do not report the service quality natively, they might have to be adapted for CASCA.

In such case, 
the client is recommended to be extended with one or more \textit{continuous reporters}. Each continuous reporter is a small module 
whose concern is sending SLO-relevant telemetry to the necessary middleware or middlewares, without understanding the semantics of such telemetry. 
Their role is to serialize this information (e.g., to JSON, YAML, CSV, ASN.1) and to send this information to the telemetry middleware. This design allows continuous reporters to be reused, e.g.,  
a continuous reporter for a specific Android client can be integrated into other Android clients for different services, as they are application-agnostic. Moreover, if 
different data serialization formats or communication protocols are needed, a single client can have multiple continuous reporters. Finally, 
continuous reporters 
must be compatible with the appropriate lifecycle management tools of the technology used to implement them (e.g., context managers, try-with-resources). 

Aside from these elements, CASCA defines the \textit{telemetry database}, which serves as a 
unified telemetry data repository. The telemetry database 
serves two purposes: providing a view of current SLO fulfilment, and storing historical data about SLO-related metrics that for analysis or 
AI training. 
CASCA is technology-agnostic for the telemetry database, but 
it is recommended to use a time-series database, given that time is a crucial aspect of telemetry and the data volume stored is expected to be large. 
Similarly, there is no constraint on the architecture of the database: while only one service is shown for simplicity, it is possible to use replicated, distributed, sharded, or other database architectures, as long as data can be accessed through a single API. 
Existing off-the-shelf database services may be leveraged for this task, such as 
InfluxDB.

Bridging the telemetry middleware and database is the role of the \textit{telemetry hook}: a service meant specifically for CASCA 
that receives a subset of the messages sent to the telemetry middleware and stores their data the telemetry database. 
A telemetry hook 
focuses on a specific type of data (e.g., temperature readings), 
deserializes these messages and converts them to a given schema for storage. Through schema conversion, 
telemetry hooks can 
store 
message metadata (e.g., sender identity) 
, as well as 
discard irrelevant fields. 
Telemetry hooks 
decouple 
the network protocol, serialization format, target database, and schema of telemetry data to maximize reuse. 
A telemetry hook is designed to target a given protocol, such as MQTT 
or HTTP. 
Each telemetry hook also contains a \textit{database connector}, which handles the interface with a specific telemetry database. At the same time, the database connector must be provided a \textit{message interpreter}, which deserializes the data received and converts it 
to a given schema for storage. These schemas are 
specified 
in external \textit{hook configuration} files. 
This allows reusing components across configurations, (e.g. each the MQTT, InfluxDB, and JSON components may be reused individually). 
Furthermore, the declarative nature of the configuration files allows for quick deployment of multiple telemetry hooks for different types of information.

CASCA receives SLO-related data 
from its 
source through 
the telemetry middleware, which 
then sends this data to its relevant telemetry hook. 
The hook deserializes the data and 
transforms it 
to the schema used for storage before storing it into 
the telemetry database.

Carbon footprint is an exception to general SLO information because of its specially challenging characteristics, as it depends not only on internal service or infrastructure data, but also on external data on the carbon intensity of energy. 
As carbon-awareness is a fundamental part of CASCA, and so is simplifying SLO management, \textit{EMMA} was designed to simplify the integration of carbon-awareness in SOA management. EMMA is a service that exposes a carbon intensity API to 
decision systems. This API allows the decision systems to retrieve the carbon intensity (i.e., the CO2 emissions per unit of energy) of a given energy mix, as well as to retrieve location-based carbon intensity data by querying governmental or predicted information. The API for EMMA should also be defined following the OpenAPI specification to enable 
drop-in replaceable versions of EMMA that can be queried from the decision system using the same SDK. 



\subsection{Service API}\label{subsec:service-api} 

The \textit{service API} is a central element in CASCA that 
fulfils two roles: it acts as an API gateway to 
the SLO fulfilment data and 
service control, and it provides a layer of privacy between the service developer 
and the CC provider. 
The service API, as represented in \Cref{fig:architecture}, is a module designed specifically for CASCA but made to be reusable, and integrates two main submodules: the service controller and the SLO controller. 

First, 
it is necessary to 
allow decision systems to reconfigure the service. Nonetheless, this service control entails considerable complexities. On the one hand, the interfaces for 
service 
reconfiguration 
vary across services, from configuration files to CLIs or well-structured APIs, 
with different communication protocols and data formats. One of the main concenrs of the service API is abstracting and unifying this interface. 
On the other hand, service operators may want to restrict how their service may be reconfigured. 
This includes denying access to specific settings, 
limiting the values for certain settings, 
or merging multiple settings into one. 
The service API thus implements a \textit{service controller} submodule to perform these tasks.

The service controller is a part of 
the service API's, such as a class or a library, that implements service control for the API. 
The only constraint in the service controller is a contract: it must allow 
the retrieval of the value of a given setting, changes to such value, 
and 
listing the existing settings and their description (e.g., name or identifier, data type, allowed values). 
Its connection to the service should be handled by lifecycle management tools, in the same manner as continuous reporters. This contract ensures that different controllers can be \textit{drop-in replaceable} for one another when services are changed. 
Thus, the service controller 
must be adapted to ensure service operators can implement the necessary service interface, ensuring compatibility. Nonetheless, for 
structured APIs, it is possible to implement general service controllers that can be then configured declaratively (e.g., an HTTP/JSON service controller).

The other submodule of the service API is the \textit{SLO controller}, which allows the service API to interact with 
SLO and telemetry data. 
The SLO controller 
retrieves and deserializes data from a given telemetry database, potentially 
including additional information relevant to the SLO represented by the data (e.g., 
name, 
description, 
fulfilment condition). 
The SLO controller performs this deserialization based on specifications for each SLO, that state how to retrieve the relevant data 
from the telemetry database, what is its data type, 
and fulfilment information. These specifications are loaded by the 
\textit{SLO parser} 
in a given data format and scheme (e.g., JSON with a given JSON schema). 
Thus, the service operator can simply write \textit{SLOs configuration} files 
in said format and schema to enable their observability. 
Moreover, a single file may contain multiple SLOs, hence the name SLOs configuration. This declarative approach simplifies adding, modifying, or removing SLOs over time. 
Furthermore, the modular design of the SLO controller makes it easy to extend, replace, and reuse elements to adapt to different telemetry databases and 
formats.

The service API, 
relying on these submodules, offers two APIs to 
decision systems. The \textit{SLO API}, which provides observability of SLOs and SLO fulfilment, and the \textit{service control API}, which enables consulting and changing the 
service configuration. 
These two APIs are offered as part of a single artifact. 
CASCA only 
imposes two requirements on these APIs: they must use network protocols that enable remote procedure calls, 
and such procedure calls must be as technology-agnostic as possible. This decouples 
the technologies used to implement the service API and 
the decision system. 
It is thus recommended to offer OpenAPI-compliant definitions for the APIs~\cite{OpenAPI} to fulfil both requirements. 

Both APIs offered by the service API should 
be reusable for different services, while supporting a minimum set of functions. 
Specifically, the SLO API must support, at least, listing all configured SLOs (i.e., SLO discovery), 
and providing the description and current value of a given SLO. 
The service control API must similarly implement configuration discovery, setting description, and setting value retrieval, but it must also implement reconfiguration to allow setting values to be changed. CASCA does not require these functionalities to be each supported by a different endpoint. 

Finally, the service API enables the service developer to privately allow an infrastructure provider 
to track SLO fulfilment and control the service. This privacy is achieved by semantically decoupling SLO data and service configuration from the information provided through the APIs. 
For example, a service operator may wish to allow a decision system to tune the number of threads assigned to video processing to reduce CPU usage below 80\% while keeping FPS between 24 and 30. 
To preserve privacy, they may declare this scenario as an \textit{A} SLO fulfilled below 80, a \textit{B} SLO fulfilled between 24 and 30, and a \textit{C} setting. Thus, 
the decision system may 
learn about SLO-configuration interplay, but not about their semantics.



\subsection{Decision system integration}\label{subsec:ai}

CASCA, following MSA principles, focuses on giving the \textit{decision system} freedom to be implemented without imposing any specific technology, library, or architecture. Thus, it is represented in \Cref{fig:architecture} as an ad-hoc developed system, as CASCA does not enforce reusability upon it. To simplify development easier, the decision system is offered 3 APIs from CASCA: the SLO API, the service control API, and the carbon footprint API, to keep track of SLOs and reconfigure the service. These APIs can be programmatically interpreted, potentially including automatic code generation, so the decision system can immediately generate an appropriate SDK for its technology. Alternatively, the API may be consumed in raw form 
(e.g., through HTTP calls). 

The freedom given to the decision system is not limited to technologies, but also includes freedom in terms of decision techniques. Traditional optimization systems, heuristics, AI-based decision systems, among others, may be implemented using CASCA. 
Furthermore, CASCA was designed to support the specific extra needs of the AI learning process to enable online training. 
As a final note, within the framework proposed in~\cite{DBLP:journals/internet/HerreraSD26}, CASCA is an MSA-oriented design to implement both the \textit{configuration API}, called service API in CASCA, and EMMA, which shares name in both proposals. 
The service API 
implements parts of both \textit{DEVOL}, through the SLO API, and \textit{POPOL}, through the service control API.




\section{Use case: Media streaming service}\label{sec:use-case-implementation}


To show the applicability of CASCA in practical environments, this section details its implementation for the control of a media streaming service. The use case is described in more detail in \Cref{subsec:use-case}, and its implementation is presented in \Cref{subsec:implementation}. Additional details on the decision systems implemented is provided in \Cref{subsec:decision-systems}.

\subsection{Use case description}\label{subsec:use-case}

The use case introduced to show the practical relevance of CASCA comes from the entertainment domain: media streaming. More specifically, following the description in~\cite{DBLP:journals/internet/HerreraSD26}, the use case assumes that a company, as a service developer, 
offers a media streaming service. This service could be compared to a small-scale 
Netflix, 
offering on-demand films to users. As such, the service has very high-quality versions of these films. However, their large size makes it impossible to directly stream films. Users may instead stream it in 
lower 
resolutions, such as 
720p. 
To enable this functionality, the media streaming can transcode 
the video file to a lower quality during runtime. 
However, transcoding becomes a key process, 
as slow transcoding may result in stuttering or buffering. Moreover, transcoding is a computationally intense process, 
and performing it requires a significant amount of resources, thus 
consuming a significant amount of energy.

To ensure users have a fluid streaming experience, a performance SLO is defined, which is fulfilled when the transcoding is performed, at least, at the same pace the video is being streamed (i.e., when FPS are above 24). In contrast, the CC provider defines an SLO that sets an upper bound on the carbon footprint of the service. These two SLOs must be traded off, as 
maximizing the performance of transcoding leads to high energy consumptions and carbon footprints. 
To make sure the infrastructure provider can reconfigure the media server with decision-making systems, CASCA is implemented to control and observe the media streaming service.

\subsection{Implementation details}\label{subsec:implementation}

\begin{figure*}
    \centering
    \includegraphics[width=0.75\textwidth]{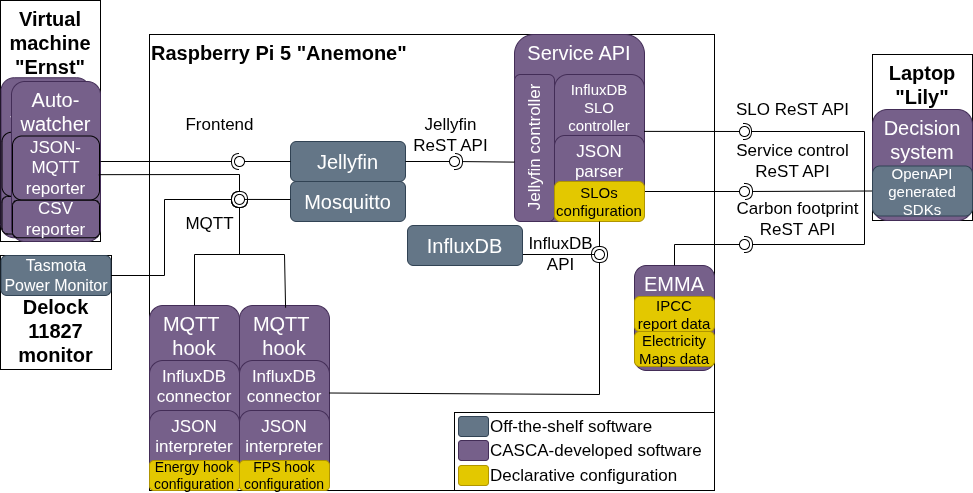}
    \caption{Implementation of CASCA for the media streaming use case.}
    \label{fig:implementation}
\end{figure*}

\Cref{fig:implementation} shows a summary of the implementation of CASCA for the media streaming use case, including deployment details that are discussed in \Cref{subsec:setup}. 
Notably, as shown by \Cref{fig:architecture}, some of these modules may be reused to implement CASCA for other use cases. The implementation aims to be simple to replicate by leveraging, whenever possible, free and open source software.

As an implementation for the media streaming service, 
Jellyfin\footnote{\url{https://jellyfin.org/}} was used, given its open source nature and widespread use. 
Jellyfin offers both a ReST API, which can be used 
to configure the service, and a web-based frontend. 
Jellyfin implements transcoding capabilities, and the configuration of this transcoding can be changed through the API. Jellyfin's frontend also has an option, Playback Info, that reports the FPS achieved during 
transcoding process 
to the user. While this information is not available at the service itself through the API, its measurement at the service side makes telemetry feasible.

To automate this process, a web client for Jellyfin, named \textit{Autowatcher}, was developed using 
Selenium\footnote{\url{https://www.selenium.dev/}}. More specifically, 
Selenium controls a browser that automatically logs into the service using a specified account, watches a specified film, and continuously reports on the content of its Playback Info readings, including FPS, simulating a user with enabled telemetry. 
It is important to note that this client uses a standalone version of Selenium, rather than Selenium Grid, to ensure accuracy to a real user environment. The Autowatcher implements a continuous reporter, which serializes the information in JSON format and uses MQTT as a telemetry protocol. Moreover, the Autowatcher also implements a CSV mass reporter, which saves all the information obtained throughout the execution of the service as a CSV file. 
This implementation shows that clients may include non-CASCA telemetry or reporting additions while still being CASCA-compliant.

MQTT was chosen as a telemetry protocol, given its lightweight nature and its design oriented to telemetry. This protocol requires a broker, which acts as the telemetry middleware element in the CASCA architecture. 
The popular open source MQTT broker Eclipse Mosquitto\footnote{\url{https://mosquitto.org/}} is used to fulfill this role. Aside from the Autowatcher instances, an additional data source sends messages to Mosquitto: a power monitoring device. More specifically, a power outlet that uses the open source Tasmota\footnote{\url{https://tasmota.github.io/}} firmware. The Tasmota Power Monitor continuously reports its readings through MQTT directly to the Mosquitto broker, 
integrating energy consumption, 
used by the CC provider's carbon footprint SLO, into CASCA. 

MQTT messages are received by two telemetry hooks that leverage MQTT, and include a JSON interpreter. One is configured to deal with energy readings from Tasmota, while the other one handles the Autowatcher's reports on FPS,  as 
depicted 
in \Cref{fig:implementation}. It is noteworthy that both hooks are the exact same software artifact, only differing in their configuration, which showcases the reusability of the CASCA architecture. These hooks send the data to the telemetry database: 
InfluxDB OSS v2\footnote{\url{https://docs.influxdata.com/influxdb/v2/}}. 
InfluxDB 
was chosen because of its design as a general-purpose time-series database. Each MQTT hook is thus equipped with an InfluxDB connector.

The service API was implemented complying with 
OpenAPI, 
using 
Python Flask and with a baseline generated with 
the open source tool OpenAPI Generator\footnote{\url{https://openapi-generator.tech/}}. 
It exposes both the SLO and service control APIs as ReST APIs to the outside. Internally, a Jellyfin controller module leverages Requests\footnote{\url{https://docs.python-requests.org/en/latest/index.html}} to interact with Jellyfin's ReST API. The SLO controller, on the other hand, queries InfluxDB. 
Its SLO parser is meant for JSON files. A single configuration file stores all the SLOs exposed by the service API, including their descrptions, and allow the service operator to describe how to get them by using an InfluxDB query. This allows a service operator who is familiar with InfluxDB to naturally apply their know-how to define CASCA SLOs, and unfamiliar operators may learn from existing InfluxDB material.

Finally, EMMA was also implemented to be 
OpenAPI-compliant, 
using 
Python Flask and 
OpenAPI generator. To obtain data on carbon intensity, EMMA has two different sources, which are bundled with it. For 
energy mixes, 
EMMA extracts data from the IPCC Climate Change Report~\cite{IPCC2014}. 
For location-related data, EMMA takes the data directly from the ElectricityMaps dataset~\cite{ElectricityMapsJul2025}. EMMA has hourly, daily, monthly, and yearly information on the carbon emissions of all countries that were, as of 2025, members of the European Union. 

In terms of provisioning, following the MSA philosophy, the implementation is fully containerized. Every independent module shown in \Cref{fig:implementation} is packaged as a Docker container, except for the Tasmota power monitor due to its embedded firmware characteristics. As all these modules interact using network protocols, they can be deployed both in a single machine and distributed into different machines, which is eased by their containerization. Finally, in accordance to open science principles and the open source nature of the implementation, all the source code, including Dockerfiles, 
Docker-Compose 
specifications, OpenAPI Generation configurations, and detailed documentation, are freely available to replicate the implementation. Specifically, EMMA\footnote{\url{https://doi.org/10.5281/zenodo.18622012}} and the Autowatcher\footnote{\url{https://doi.org/10.5281/zenodo.18622399}} are provided as separate repositories, while the rest of the modules are provided as a single repository\footnote{\url{https://doi.org/10.5281/zenodo.18622182}} to streamline their deployment. It is noteworthy that data or software that belongs to third parties may not be included in these repositories to comply with copyright laws and licensing terms, requiring interested readers to provide their own copies.

\subsection{Decision systems}\label{subsec:decision-systems}

\begin{table}[!h]
\centering
\caption{List of notations.}
\begin{tabular}{cp{0.7\columnwidth}}
\hline
\textbf{Parameter} & \multicolumn{1}{c}{\textbf{Meaning}}                                         \\ \hline
\multicolumn{2}{c}{\textbf{Common notations}}\\ \hline
$t$                & Time-step                                            \\
$\tau$                & Duration of a single time-step                                       \\
$P$                & Set of all the configuration parameters of the service                                                  \\
$p$                & Specific parameter of the service, $p\in P$\\
$p_{\min}$, $p_{\max}$           & Minimum and maximum values that can be given to $p$, i.e., $p\in[p_{\min}, p_{\max}]$ \\
$S$         & Set of all SLOs of the service \\
$s$          & Specific SLO, $s\in S$ \\
$s_{\min}$, $s_{\max}$          & Minimum and maximum values for $s$ to be fulfilled. If $s\notin[s_{\min}, s_{\max}]$, s is considered to be unfulfilled\\
$s(t)$, $p(t)$                & Values of $s$ and $p$, respectively, at time-step $t$\\

\hline
\multicolumn{2}{c}{\textbf{RDS notations}}\\ \hline
$n(t)$         & Random number generated at time-step $t$ so that $n(t)\in[p_{\min}, p_{\max}]$ \\

\hline
\multicolumn{2}{c}{\textbf{GDS notations}}\\ \hline
$\Delta$                & Maximum rate of change for $p$ in a single time-step                                           \\
$\Lambda$                & Correlation of $p$ and $s$. $\Lambda=1$ if they are directly correlated, and $\Lambda=-1$ if they are inversely correlated \\
$isC$                & Boolean value that is true if $s$ represents the carbon footprint\\
$API_{SLO}$         & CASCA SLO ReST API \\
$API_{CONT}$        & CASCA service control ReST API \\
$API_{EMMA}$        & CASCA carbon footprint ReST API \\

\hline
\multicolumn{2}{c}{\textbf{RLDS notations}}\\ \hline
$S_R$                       & Set of SLOs selected to be considered by RLDS, $S_R\subseteq S$ \\
$P_R$                       & Set of parameters selected to be controlled by RLDS, $P_R\subseteq P$\\
$C$                         & Carbon footprint SLO, mandatory to use RLDS and cannot be de-selected, $C\notin S_R$\\
$C(t)$                         & Carbon footprint at time-step $t$\\
$\mathcal{S}$                & State space \\
$\sigma(t)$                 & State at time-step $t$, $\sigma(t)\in \mathcal{S}$\\
$\mathcal{A}$               & Action space \\
$\alpha(t)$                 & Action taken by RLDS at time-step $t$, $\alpha(t)\in \mathcal{A}$\\
$\mathcal{R}$               & Reward function \\
$\mathcal{R}(t)$            & Reward given to RLDS at time-step $t$ \\
$IN(x, a, b, c)$            & Auxiliary function, takes a value of $1$ if $x\in[a, b]$ and $-2c$ otherwise\\
$\psi(\mathcal{A}|\mathcal{S})$ & Reinforcement Learning policy, $\alpha(t) = \psi(\sigma(t))$ \\
$\Psi(\mathcal{A}|\mathcal{S})$ & Optimal policy, guarantees the optimal action for any state in the state space \\
\hline
\end{tabular}
\label{tab:notations}
\end{table}

The decision systems embody their homonyms in CASCA's architecture, and include both AI-based and non-AI based decision systems. 
These use case implementation use SDKs for the service APIs and EMMA, generated with OpenAPI generator, as well as use direct HTTP calls. 
To show the flexibility of this approach in terms of used technologies, three decision systems have been developed, each leveraging a different programming language: Bash, Rust, and Python, which are presented in the remainder of this section in ascending order of complexity. The notation used throughout this section is detailed in \Cref{tab:notations}. For the common notations, $\tau$ represents the duration of a time-step, while one decision is taken per time-step $t$, $P$ represents all configuration parameters, $p\in P$ represents a specific parameter, $p_{\min}$ and $p_{\max}$ represent the minimum and maximum value that can be given to $p$, $S$ represents all SLOs, and $s\in S$ a specific SLO. The range in which $s$ is considered to be fulfilled is denoted by $s_{\min}$ and $s_{\max}$, and time, expressed as $t$, is expressed in sequential time-steps of length $\tau$.

\subsubsection{Random Decision System}\label{subsubsec:random}

The first decision system represents the very simplistic baseline of taking random decisions, and is thus named the \textit{random decision system} (RDS). RDS takes as an input a configuration parameter $p$, the time-step length $\tau$, and the range of $p$, i.e., $p_{\min}$ and $p_{\max}$. RDS operates by generating a random number $n(t)\in[p_{\min}, p_{\max}]$ in each time-step, setting the configuration parameter to said value ($p(t+1) := n(t)$), and waiting for $\tau$, after which a new value ($n(t+1)$) is generated. 
As the behaviour of RDS is simple, we believe no additional pseudocode or description needs to be included. Finally, the open-source implementation of RDS\footnote{\url{https://doi.org/10.5281/zenodo.18622215}} is implemented using the Bash scripting language, and can be containerized as the rest of the architecture.

\subsubsection{Greedy Decision System}\label{subsubsec:greedy}

The second decision system, named the \textit{greedy decision system} (GDS), leverages a greedy heuristic for decision-making. This system considers one SLO, $s\in S$, and one parameter, $p$. Two additional parameters must be provided to GDS: the amount in which $p$ will be changed, denoted $\Delta$, and whether $p$ is directly or inversely correlated with $s$, denoted $\Lambda$, so that $\Lambda=1$ if they are directly correlated, and $\Lambda=-1$ if they are inversely correlated. 
Furthermore, it is possible to include an optional parameter, $isC$, that denotes if the SLO should be compounded with EMMA.

The behaviour of the greedy heuristic used by GDS in each time-step is represented in \Cref{alg:greedy}. In the time-step $t$, GDS obtains the fulfilment range for $s$, $s_{\min}$ and $s_{\max}$ (line 1), and checks if the value of $s$, denoted $s(t)$, is within range (lines 4, 8-12). If $s(t) > s_{\max}$,  GDS adjusts $p(t)$ so that $p(t+1) := p(t)-\Lambda\Delta$ (lines 8-9). If $s(t) <s_{\min}$, the adjustment is the opposite. 
(lines 10-12). If $s(t)$ is fulfilled 
or both amounts are equal, no action is taken. If the adjustment is out of the range of $p$, gathered in line 3, it is reset to $p_{\min}$ or $p_{\max}$ accordingly (lines 13-17). Finally, the change to the value of $p$ is applied so it is effective at $t+1$ (line 18) and GDS waits for the next time-step (line 19). The GDS has been implemented in Rust\footnote{\url{https://doi.org/10.5281/zenodo.18622384}}, and can also be containerized.

\begin{algorithm}[tbp]
\caption{Pseudocode for a time-step of GDS.}
\small
\label{alg:greedy}
\begin{algorithmic}[1]
\Function{GDS}{$s$, $isC$, $p$, $API_{SLO}$, $API_{CONT}$, $API_{EMMA}$, $\Delta$, $\Lambda$, $\tau$}
\State $s_{\max}, s_{\min} := API_{SLO}.describe(s)$
\State $p_{\max}, p_{\min} := API_{CONT}.describe(p)$
\State $s(t) := API_{SLO}.value(s)$
\State $p(t) := API_{CONT}.value(p)$
\If{$isC$}
    \State $s(t) := s(t) \cdot API_{EMMA}.intensity()$
\EndIf
\If{$s(t) > s_{\max}$}
    \State $p(t+1) := p(t) - \Lambda\Delta$
\Else
    \If{$s(t) < s_{\min}$}
        \State $p(t+1) := p(t) + \Lambda\Delta$
    \EndIf
\EndIf
\If{$p(t+1) > p_{\max}$}
    \State $p(t+1) := p_{\max}$
\Else
    \If{$p(t+1) < p_{\min}$}
        \State $p(t+1) := p_{\min}$
    \EndIf
\EndIf
\State $API_{CONT}.modify(p, p(t+1))$
\State $wait(\tau)$
\EndFunction
\end{algorithmic}
\end{algorithm}

\subsubsection{Reinforcement Learning Decision System}

Finally, as a representative of AI-based decision systems, a \textit{Reinforcement Learning Decision System} (RLDS), based on DRL technology, has also been implemented. 
In each time-step, RLDS gets some information about the environment, makes a decision, and obtains feedback on the effect of its decision on the environment. In CASCA, the environment is the service being controlled by RLDS, which is characterized as a Markov Decision Process (MDP). An MDP is characterized by a state space $\mathcal{S}$, which represent every possible state of the environment, an action space, $\mathcal{A}$, which are the complete set of decisions the agent can potentially make, and a reward function $\mathcal{R}(t)$, which is the feedback provided to the model. Within the context of CASCA, the state of the MDP at time-step $t$, denoted $\sigma(t)\in\mathcal{S}$, contains the values of all the configuration parameters and SLOs selected to be considered ($P_R\subseteq P$ and $S_R\subseteq S$, respectively) at time-step $t$. 
It is important that, like in GDS, the carbon footprint is a special case. In RLDS, the carbon footprint SLO, $C$, already contains EMMA information and is mandatory, and thus outside the SLOs that are selected to be considered. 
Formally, a state is defined by \Cref{eq:state}.

\small
\begin{equation}
\begin{split}
    \sigma(t) = \{p_{\min}, p_{\max}, p(t) | p\in P_R\}\cup \{s_{\min}, s_{\max}, s(t)|s\in S_R\}\\\cup \{C(t)\}
    \label{eq:state}
\end{split}
\end{equation}
\normalsize

Based on this state, RLDS takes an action at time-step $t$, denoted $\alpha(t)$. An action in RLDS is a new configuration for the service, i.e., a value for every parameter within its allowed range, as per \Cref{eq:action}. After taking such action, the RLDS agent waits for $\tau$ to ensure the effects of the action are reflected in the SLOs. It then receives a reward, based on the number of SLOs fulfilled and the carbon footprint at $t$, denoted $C(t)$. Each fulfilled SLO adds $+C(t)^{-1}$ to the reward, and each unfulfilled SLO adds $-C(t)$ instead. This reward function thus positively rewards fulfilled SLOs, and negatively rewards unfulfilled SLOs. At the same time, fulfilled SLOs can obtain a higher reward by reducing their carbon footprint, while unfulfilled SLOs can reduce the negativity of their reward by reducing their carbon footprint. The final sum of these contributions 
is 
then normalized. 
To mathematically define the reward function, let $IN(x, a, b, c)$ be a function that takes a value of $1$ if $x\in[a, b]$, and a value of $-2c$ if $a>x$ or $b<x$. Then, the reward function is defined by \Cref{eq:reward}. Once the reward is received, RLDS retrieves a new state ($\sigma(t+1)$) and restarting the loop.

\small
\begin{equation}
    \alpha(t) = \{a: a\in[p_{\min}, p_{\max}] | p\in P_R\}
    \label{eq:action}
\end{equation}

\begin{equation}
\begin{split}
    &\mathcal{R}(t) = \frac{1}{|S_R|}\\&\sum_{s\in S_R}\frac{IN(s(t+1), s_{\min}, s_{\max}, C(t+1))}{C(t+1)}
     \label{eq:reward}
\end{split}
\end{equation}

\normalsize

In Reinforcement Learning, an is said to follow a \textit{policy} $\psi(\mathcal{A}|\mathcal{S})$, which is a function that defines, for each state $\sigma(t)$, which is the action $\alpha(t)$ that the action will take. Thus, the behaviour of an agent in a time-step with a set policy can be summarized as $\alpha(t) = \psi(\sigma(t))$. There is also an especially interesting policy, the optimal policy, denoted $\Psi(\mathcal{A}|\mathcal{S})$. 
The objective of a DRL is to learn from the environment to 
approximate the optimal policy. In the case of RLDS, the agent uses a technique known as Proximal Policy Optimization (PPO), which tries to directly learn a policy. 
RLDS has been implemented using Python\footnote{\url{https://doi.org/10.5281/zenodo.18622320}}.
\section{Evaluation}\label{sec:evaluation}

To assess the feasibility of using CASCA, this section presents an evaluation over the media streaming use case, deployed in a local testbed. \Cref{subsec:setup} presents the setup used for evaluation, while the obtained results are analyzed and discussed in \Cref{subsec:results}.

\subsection{Evaluation setup}\label{subsec:setup}


The evaluation makes use of the implementation detailed in \Cref{sec:use-case-implementation}, including the software, Dockerfiles, and Docker Compose specifications that are linked in it. The deployment has been distributed into a real testbed with 3 machines: a Raspberry Pi 5 Model B Rev. 1.0, a virtual machine, and a development laptop, as shown in \Cref{fig:implementation}. These machines are labelled by their hostnames, Anemone, Ernst, and Lily, respectively, to make them easier to reference throughout the text. Anemone has a 4-core 64-bit ARM CPU, 8 GB RAM, and runs Raspbian Bookworm. Ernst has a 16-core 64-bit x86 CPU, 8 GB RAM, and runs Arch Linux. Lily has a 12-core 64-bit x64 CPU, 32 GB RAM, and also runs Arch Linux.

In terms of software, all the components of the CASCA implementation, except for the Tasmota Power Monitor, are deployed in Docker containers. The Tasmota Power Monitor, as an embedded system, is executed by a Delock 11827 monitor, which Anemone is plugged into to get energy readings. Anemone is monitored for energy as it hosts the services in CASCA that would be executed at the server-side. These services are Jellyfin, Mosquitto, the MQTT hooks, InfluxDB, and the service API. EMMA is also executed in Anemone to simplify the network structure, although it may be executed in any of the three machines. Ernst, instead, hosts the client-side part: autowatchers. Specifically, Ernst executes two autowatchers to simulate service load, both of which are set to watch the public domain movie \textit{Night of the Living Dead}. This number of autowatchers was determined empirically, considering Anemone's capabilities. Finally, Lily executes the decision systems in a separate environment, like infrastructure providers would.

Network-wise, Anemone only exposes to the outside its frontend, MQTT broker, and the three ReST APIs consumed by the decision systems. Moreover, the Delock power monitor, Anemone, and Ernst are all connected to an internal network, Ernst through Ethernet, and the other two via Wi-Fi. Lily is connected to a separate network via Ethernet, and only has access to the ReST APIs. For the network overhead comparison experiments, Lily has also been given direct access to the Jellyfin and InfluxDB APIs to ensure a fair comparison. Carbon-intensity data was gathered through EMMA at the most precise granularity available (i.e., hourly) for the appropriate timestamps and Austria, as Anemone is located within Austria.

The three decisions systems proposed, namely RDS, GDS, and RLDS, are used in the evaluation, and are implemented as described in \Cref{subsec:decision-systems}. In all cases, the decision systems are set to control the variable named \textit{Encoding Thread Count}, with a range of $[p_{\min}, p_{\max}]=[0, 16]$, which governs the number of threads used by the media server to perform transcoding. To ensure that this setting takes effect, the media server is configured to continuously transcode the video from its original 4K quality to 720p. Two SLOs are declared: transcoding FPS, referred to simply as \textit{FPS}, and Anemone's apparent power consumption, denoted simply as power consumption. Both SLOs are computed continuously as a rolling mean during the last 5 minutes, and FPS have a range of $[s_{\min}, s_{\max}]=[24, 30]$, as per the frame rate of the video used. GDS is configured to use FPS as its SLO, with $\Lambda=1, \Delta=1$, while RLDS uses both the FPS as its main SLO, and the power consumption to calculate the carbon footprint through EMMA, as per its reward function. RLDS results are obtained from its training stage, as it is used to represent the ability of decision systems to learn. In the experiment with obscured parameters, \textit{Encoding Thread Count} is renamed to \textit{ServiceParam}, and \textit{FPS} is renamed to \textit{ServiceSLO}, providing no information of their underlying parameters and SLOs. The power consumption, while also offered as an SLO, was not obscured, as it is measured by the CC provider, not by the application developer.

The evaluation has four objectives: to validate the technology-agnostic aspects of CASCA, to measure the effects of privacy in the agents' performance, to assess its overhead, and to evaluate its flexibility for reconfigurations. As it is complex to validate the first objective directly, the evaluation assesses and compares the performance of the three proposed agents, RDS, GDS, and RLDS.


\subsection{Evaluation results}\label{subsec:results}


\begin{figure*}
    \begin{minipage}{0.65\columnwidth}
    \captionsetup{type=figure}
    \includegraphics[width=\columnwidth]{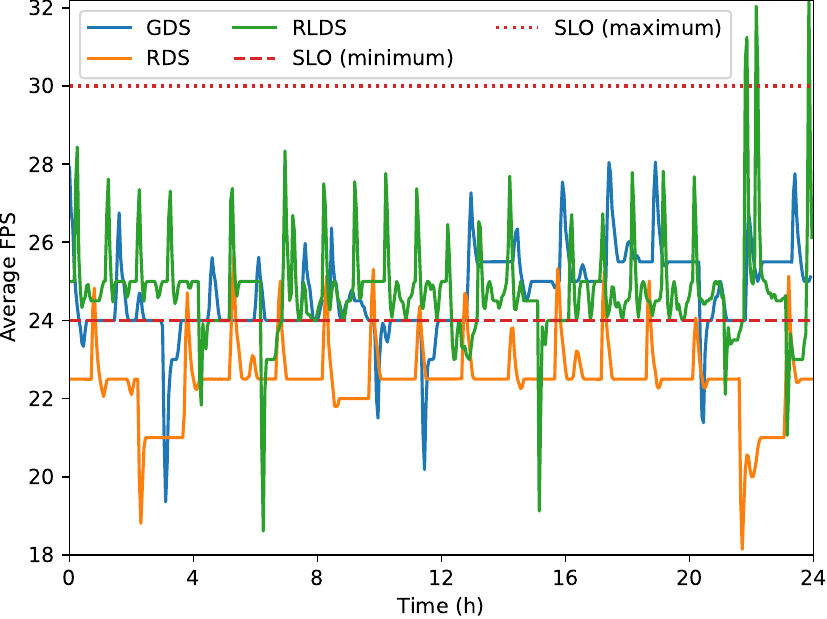}
    \caption{Average frames per second achieved over time with all decision systems and full descriptions.}
    \label{fig:fps-revealed}
    \end{minipage}\hfill
    \begin{minipage}{0.65\columnwidth}
    \captionsetup{type=figure}
    \includegraphics[width=\columnwidth]{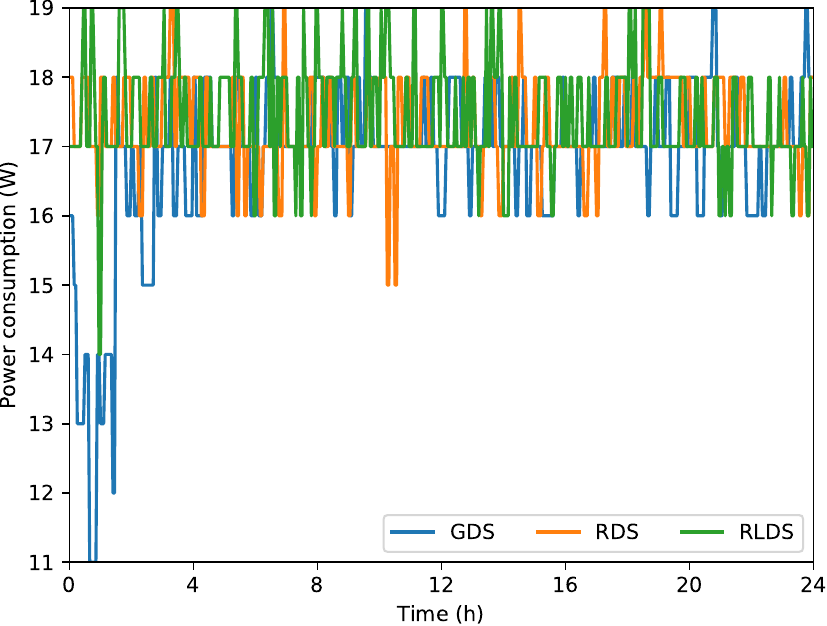}
    \caption{Power consumption over time with all decision systems and full descriptions.}
    \label{fig:power-revealed}
    \end{minipage}\hfill
    \begin{minipage}{0.65\columnwidth}
    \captionsetup{type=figure}
    \includegraphics[width=\columnwidth]{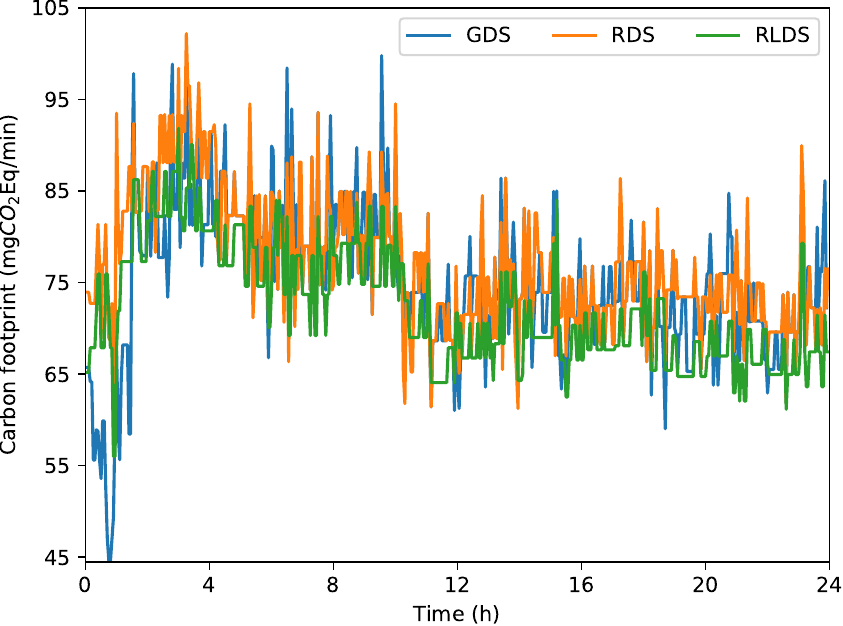}
    \caption{Carbon footprint over time with all decision systems and full descriptions.}
    \label{fig:carbon-revealed}
    \end{minipage}
\end{figure*}

The first experiment assesses the performance of the decision systems when both the FPS SLO and the EncodingThreadCount parameter are appropriately named and described. The performance is analyzed with three separate metrics: frames per second, power consumption, and carbon footprint. All these experiments, that serve as a baseline, were executed during 24 hours for each decision system. Starting with the performance in terms of frames per second, depicted by \Cref{fig:fps-revealed}, the three decision systems show different behaviour. In all cases, the trends exhibit clear ascents and declines among a mostly stable behaviour, which are generated by the clients watching the movie. Descents occur as a result of a client starting to watch the video once it is finished. At this stage, some buffering occurs, in which FPS lower until enough video is transcoded to start serving it to the client. This buffering negatively impacts the average FPS gathered per minute, and the longer the buffering takes, the lower the descent. Similarly, ascents occur when the media server provides cached transcoded video. In this figure, the SLO is considered fulfilled if its value is within the region indicated by the dashed and dotted red lines, i.e., between 24 and 30. GDS is able to fulfill the SLO 90.625\% of the time (21h45min), while RLDS is close, fulfilling it 85.208\% (20h27min). Due to its random behaviour, RDS fails to set an appropriate number of threads, and it only fulfills the SLO 5.417\% of the time (1h18min). Over the experiment, RLDS achieves the best performance, with an average of 24.725 frames per second, but also the least consistent, with a standard deviation of 1.260. GDS is a close second with 24.655 frames per second, and the most consistent of the systems (0.908 standard deviation). Finally, RDS achieves the worst performance of the three at 22.468 frames per second, and a standard deviation of 1.068.

For power consumption, shown in \Cref{fig:power-revealed}, the results are similar, with GDS being the most power efficient, consuming 16.825W on average, but also the most varying, at 1.208 standard deviation. RLDS and RDS consume similar amounts of energy on average, 17.286 W for RDS (0.690 standard deviation) and 17.431 W for RDS (0.7969 standard deviation). The GDS effects of low power consumption and high variance, nonetheless, are due to a decreased power consumption at the beginning of the experiment. Interestingly, when carbon footprint is considered instead, the trend is slightly different. As \Cref{fig:carbon-revealed} shows, GDS achieves a smaller carbon footprint, despite the higher average energy consumption. On average, if RDS is used, the service emits 76.945 mg of CO2-equivalent per minute (7.270 standard deviation), with the next most carbon-intense decisions being taken by GDS, at 74.883 mg (8.260 standard deviation). The most carbon-efficient decisions, in which the service emits on average 72.377 mg of CO2-equivalent per minute, with a standard deviation of 6.574, is RLDS. This is due to the fact that RLDS is the only one of the decision systems analyzed that considers the carbon footprint and aims to minimize energy consumption as the carbon footprint rises, while GDS uses only the SLO, and RDS considers neither.

\begin{figure*}
    \begin{minipage}{0.65\columnwidth}
    \captionsetup{type=figure}
    \includegraphics[width=\columnwidth]{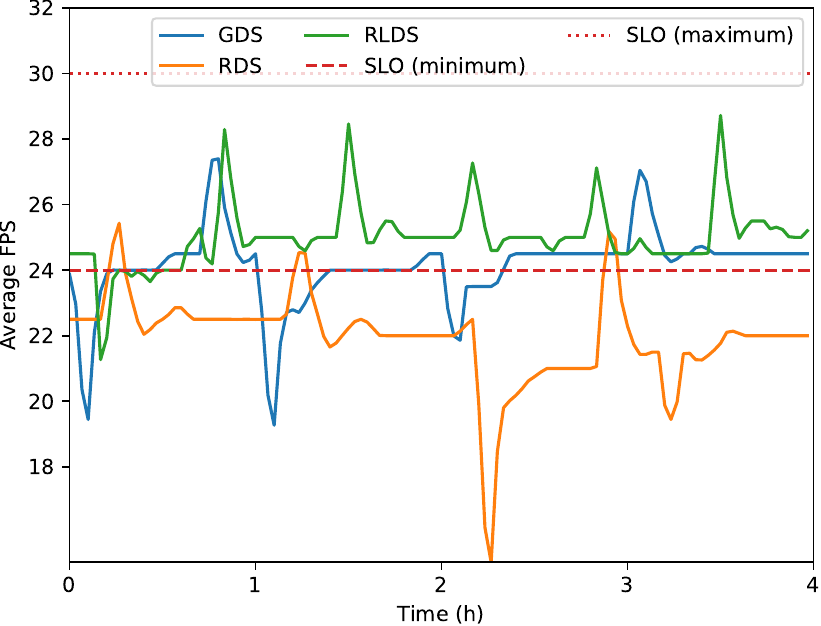}
    \caption{Average frames per second achieved over time with all decision systems and obscured descriptions.}
    \label{fig:fps-redacted}
    \end{minipage}\hfill
    \begin{minipage}{0.65\columnwidth}
    \captionsetup{type=figure}
    \includegraphics[width=\columnwidth]{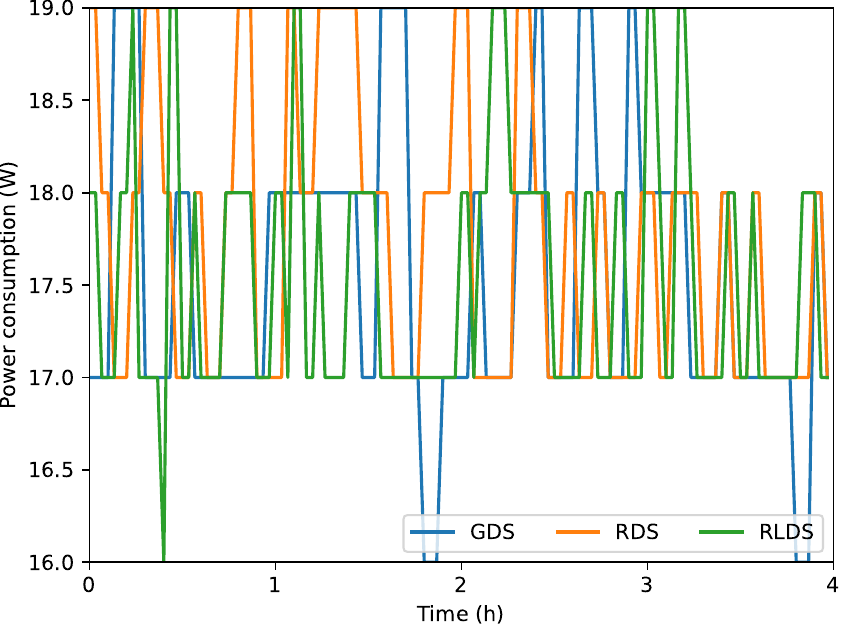}
    \caption{Power consumption over time with all decision systems and obscured descriptions.}
    \label{fig:power-redacted}
    \end{minipage}\hfill
    \begin{minipage}{0.65\columnwidth}
    \captionsetup{type=figure}
    \includegraphics[width=\columnwidth]{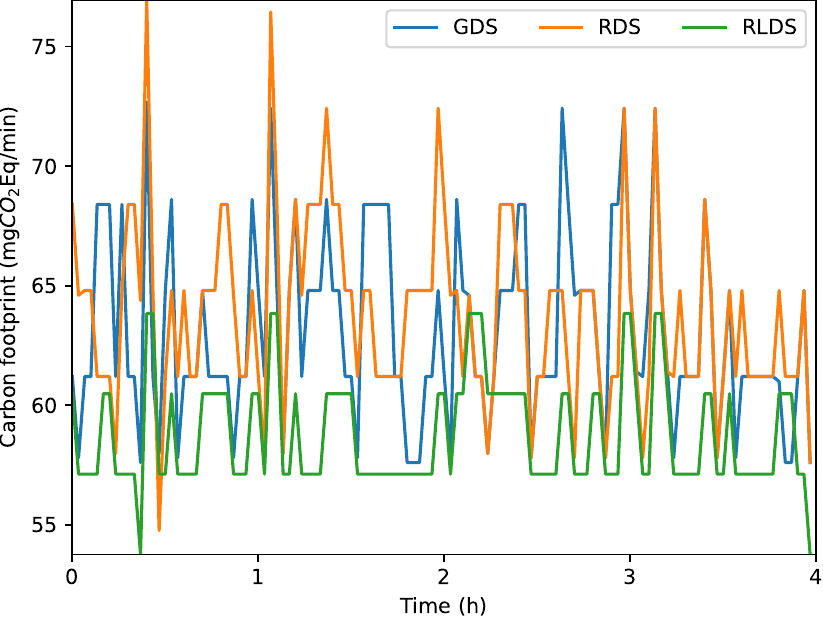}
    \caption{Carbon footprint over time with all decision systems and obscured descriptions.}
    \label{fig:carbon-redacted}
    \end{minipage}
\end{figure*}

The second experiment is aimed at assessing CASCA's ability to allow decision systems to learn and manage the service's configuration while protecting the privacy of the service. The analyses focus on the similarity of behaviour between decision systems with revealed and obscured descriptions, and hence, the experiments have been carried out for a shorter time horizon of 4 hours, enough to assess these similarities. \Cref{fig:fps-redacted} shows the results in terms of FPS, where decision systems behave similarly to \Cref{fig:fps-revealed}. GDS can fulfil the SLO, but stays relatively close to its lower limit, while RLDS fulfils it most of the time, except for a low descent in the first few minutes, and RDS only fulfils the SLO in its highest ascents. The statistics are similar: RLDS achieves the best performance with an average of 25.007 FPS (0.981 standard deviation), followed by GDS with 24.108 FPS on average (1.197 standard deviation), while RDS achieves 21.962 FPS (standard deviation 1.371). As the timeframe is shorter, ascents and descents are more significant to the overall SLO fulfilment, with RLDS fulfilling it 92.5\% of the experiment, GDS fulfilling it 75.833\% of it, and RDS 5.000\%. Power consumption (\Cref{fig:power-redacted}) is also similar to the experiment with revealed descriptions, with all three consuming similar amounts: 17.533 W for RLDS (standard deviation 0.685), 17.542 W for GDS (standard deviation 0.798), and 17.742 W for RDS (standard deviation 0.761). In the case of carbon footprint, depicted in \Cref{fig:carbon-redacted}, the decision systems behave similarly to \Cref{fig:carbon-revealed}. RLDS is the most carbon-efficient, emitting an average of 58.800mg of CO2-equivalent per minute (2.305 standard deviation), while GDS emits 63.085mg (3.928 standard deviation), and RDS is the most carbon-intense with 63.803 mg (3.847 standard deviation). Overall, all three systems behave similarly when managing interpretable parameters and SLOs (FPS and EncodingThreadCount), and obscured ones (ServiceSLO and ServiceParam), and hence, CC providers can manage services effectively while protecting the privacy of service developers.

\begin{figure*}
    \begin{minipage}{0.65\columnwidth}
    \captionsetup{type=figure}
    \includegraphics[width=\columnwidth]{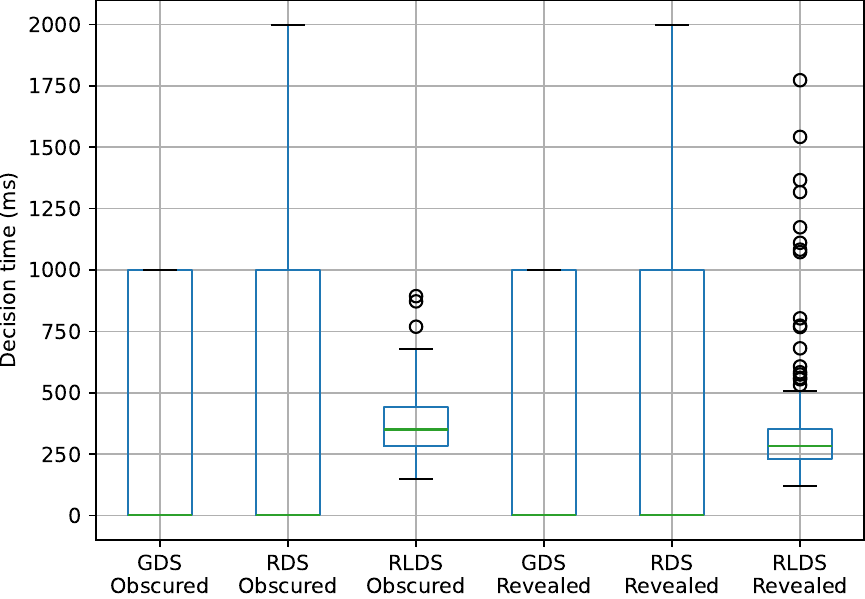}
    \caption{Time taken to make a decision by all methods in both experiments}
    \label{fig:decision-timing}
    \end{minipage}\hfill
    \begin{minipage}{0.65\columnwidth}
    \captionsetup{type=figure}
    \includegraphics[width=\columnwidth]{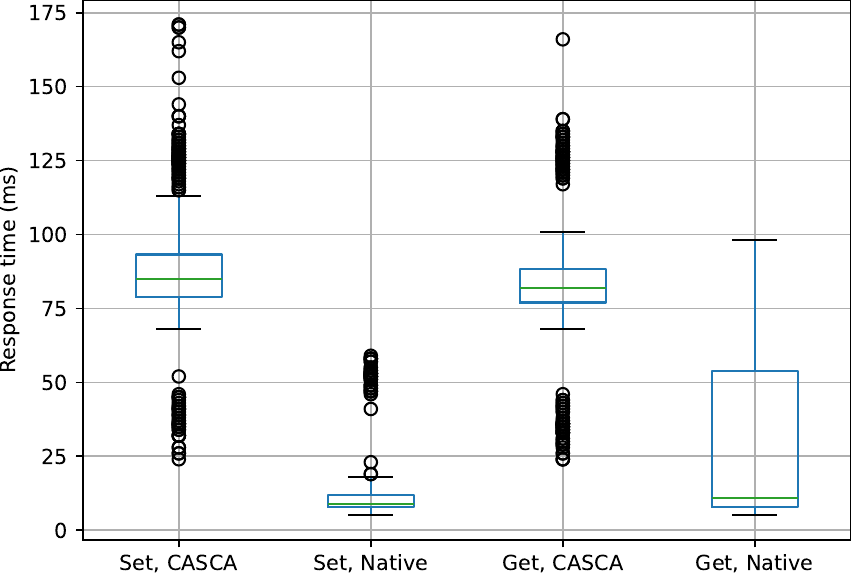}
    \caption{Response times getting and setting the value of a configuration parameter natively and with CASCA.}
    \label{fig:config-response-times}
    \end{minipage}\hfill
    \begin{minipage}{0.65\columnwidth}
    \captionsetup{type=figure}
    \includegraphics[width=\columnwidth]{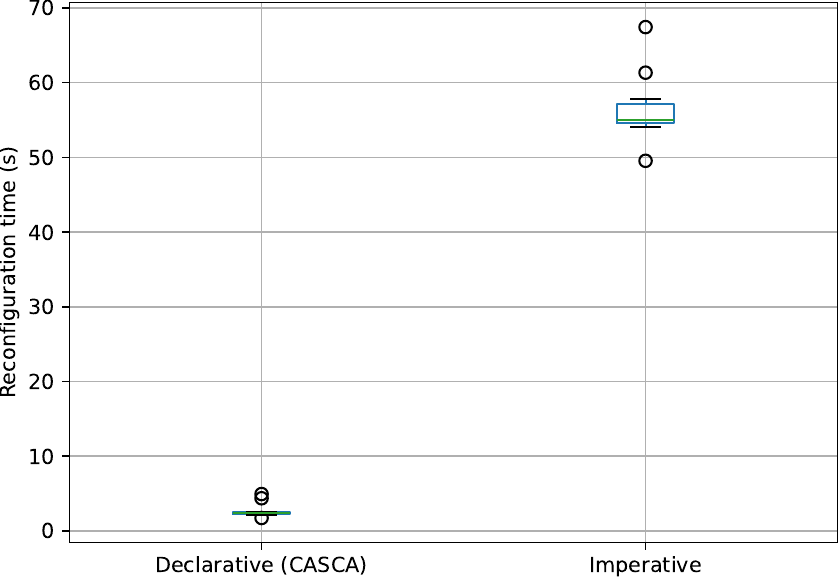}
    \caption{Reconfiguration time for modifying the configuration parameters or SLOs of the platform, with CASCA and with a traditional imperative approach.}
    \label{fig:reconfiguration-times}
    \end{minipage}
\end{figure*}
    
The next analysis instead focuses on the time taken to make a decision by the three decision systems, with both revealed and obscured descriptions. The results are shown as a series of boxplots in \Cref{fig:decision-timing}. From left to right, operating on obscured descriptions, GDS has a significant variance, requiring an average of 361.35 ms to make a decision (482.42 standard deviation), requiring up to 10000 ms in the worst case. RDS is slower, requiring an average of 445.38 ms (531.99 standard deviation), with a worst-case scenario of 2000 ms. RLDS is in the middle of the two, with an average of 376.79 ms, but the most consistent (143.26 ms standard deviation), and takes 893.45 ms in the worst case. If revealed descriptions are used instead, the results are similar: GDS and RDS have the same worst case, and their averages are of 307.53 ms (461.95 standard deviation) and 341.00 ms (478.94 ms standard deviation), respectively. RLDS is again in the middle, with an average of 313.43 ms, and is again the most consistent (165.35 standard deviation), with a worst-case decision time of 1773.223. In all cases, the results are similar across obscured and revealed values, with some outliers being higher with the revealed descriptions. As such, the effect of obscuring descriptions appears to be negligible.

To continue the timing analyses, it is necessary to evaluate the overhead of CASCA and compare it to native interaction with the service, in terms of response time. To assess this comparison for configuration parameters, the encoding thread count has been set, as well as obtained, 400 times, for a total of 1600 samples considering both setting and getting and both CASCA and native interaction. The experiments are also evenly split across the values set and got, performing 20 experiments for each value between 1 and 8. The results of this experiment are shown in \Cref{fig:config-response-times}. In terms of setting values, CASCA introduces not only an overhead, but also a higher standard deviation compared to native interactions. On average, CASCA requires 88.66 ms (23.36 standard deviation) to change a configuration parameter, while native interactions can perform these changes in, on average, 13.92 ms (12.87 ms standard deviation). If the value for a parameter is to be retrieved, instead, CASCA takes 83.115 ms on average (25.22 standard deviation), while the native interaction requires 28.67 ms on average, although it is inconsistent (24.31 standard deviation). The worst cases for CASCA are 171 ms for setting a value and 166 for getting it, while 59 ms and 98 ms, respectively, for native interactions. Overall, CASCA adds approximately 74 ms of delay for setting a value, and 71 ms for getting it. 
Overall, the overhead added by CASCA, under 75 ms, allows for a decision system to obtain the current value of a parameter and change it in approximately 167 ms. These values are lower than the standard deviations observed in \Cref{fig:decision-timing}, being similar to the standard deviation of RLDS. Thus, the overhead can be considered acceptable for the tested decision systems.

Finally, to show the flexibility of CASCA, the final experiment assesses the time needed to reconfigure an instance of the platform. A \textit{reconfiguration} in this context refers to a change in the service API that modifies the context given to the decision system: adding, removing, or changing configuration parameters or SLOs. Changes may include their names, descriptions, or types. While the configuration parameters of a service are often stable, the SLOs may vary more often (e.g., if the service implements an additional premium pricing, it may implement a new, stricter performance SLO for premium users). Hence, this experiment was performed by removing, adding, and changing SLOs 10 times for each approach. Traditionally, these changes are performed imperatively, i.e., by changing the source code of the API that refers to the parameter or SLO. In contrast, CASCA has a declarative model, where the configuration file for the parameters or SLOs can be modified and refreshed without modifying the source code. \Cref{fig:reconfiguration-times} shows the time needed to apply a reconfiguration using both approaches. CASCA takes, on average, 2.764s to apply a new reconfiguration (median 2.376s, standard deviation 1.038s). In contrast, 56.476s are required on average by the imperative approach (55.010s median, 4.850s standard deviation), representing an average overhead of 53.712s compared to CASCA. In the implementation used for evaluation, the main difference is that source code modifications require the service API's Docker image to be rebuilt, and the container stopped, destroyed, and replaced by a new one. In contrast, a change to the configuration file can be applied by restarting the existing container. Similarly, other implementations may incur in additional times if the changes require the source code to be recompiled.
\section{Conclusion and future works}\label{sec:conclusion}

Modern software, architected as an MSA, must be dynamically configured to fulfil its SLOs, providing a balance between resource consumption, including energy consumption, and service quality. The growth in interest on sustainability makes carbon-awareness a key SLO to fulfil. However, CC providers, with most knowledge and control over the system, cannot configure these services without breaking their confidentiality to the service developers. Moreover, existing control systems, especially those meant for AI, are deeply coupled to their services, counter to the MSA principles. CASCA is thus presented as an MSA-based, carbon-aware platform that enables decision systems, both AI-based and non-AI based, to reconfigure services and fulfil their SLOs, while protecting the privacy of service developers. Our results show that CASCA is quick to reconfigure, supports multiple decision systems in different technologies, and can effectively allow privacy-protecting SLO fulfilment.

Future research lines head in three different paths. On the one hand, we plan to extend CASCA with hardware control systems, enabling it to not only reconfigure the properties of the service, but also of the hardware it runs on, and thus making use of hardware-bound energy saving measures. On the other hand, CASCA is to be extended to consider service orchestration throughout the complete CC, also being able to decide where each service runs and migrating services over time. Finally, we expect to develop a dashboard service to ease the generation of CASCA declarative configurations.


\bibliographystyle{IEEEtran}
\bibliography{biblio}

\vspace{-1.5cm}

\begin{IEEEbiography}[{\includegraphics[width=1in,height=1.25in,clip,keepaspectratio]{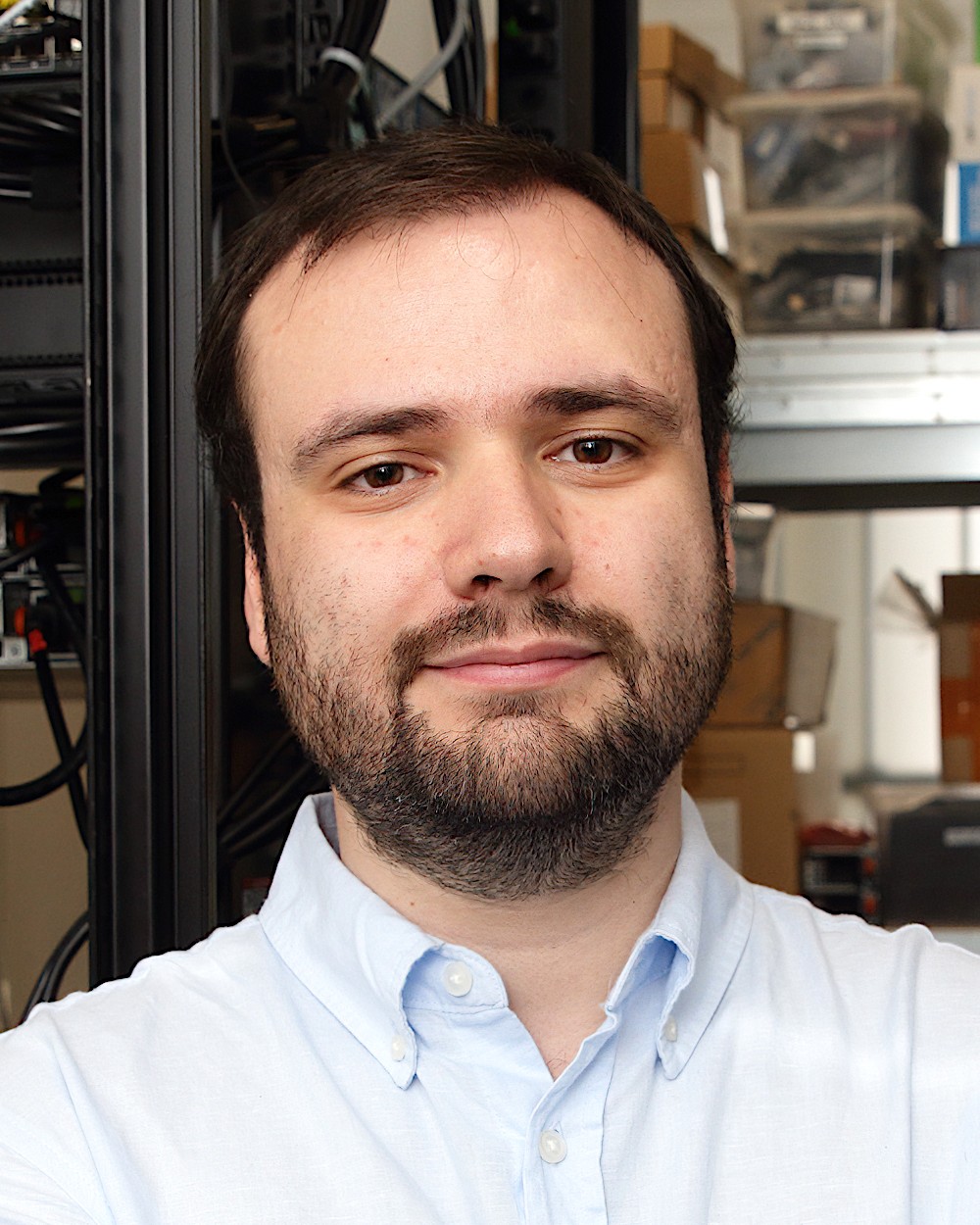}}]{Juan Luis Herrera} is a MSCA postdoctoral fellow in the Distributed Systems Group at TU Wien. He received a Ph.D. in Computer Science from the University of Extremadura in 2023. His main research interests include IoT, CC, service management, and sustainability.
\end{IEEEbiography}

\vspace{-1.5cm}

\begin{IEEEbiography}[{\includegraphics[width=1in,height=1.25in,clip,keepaspectratio]{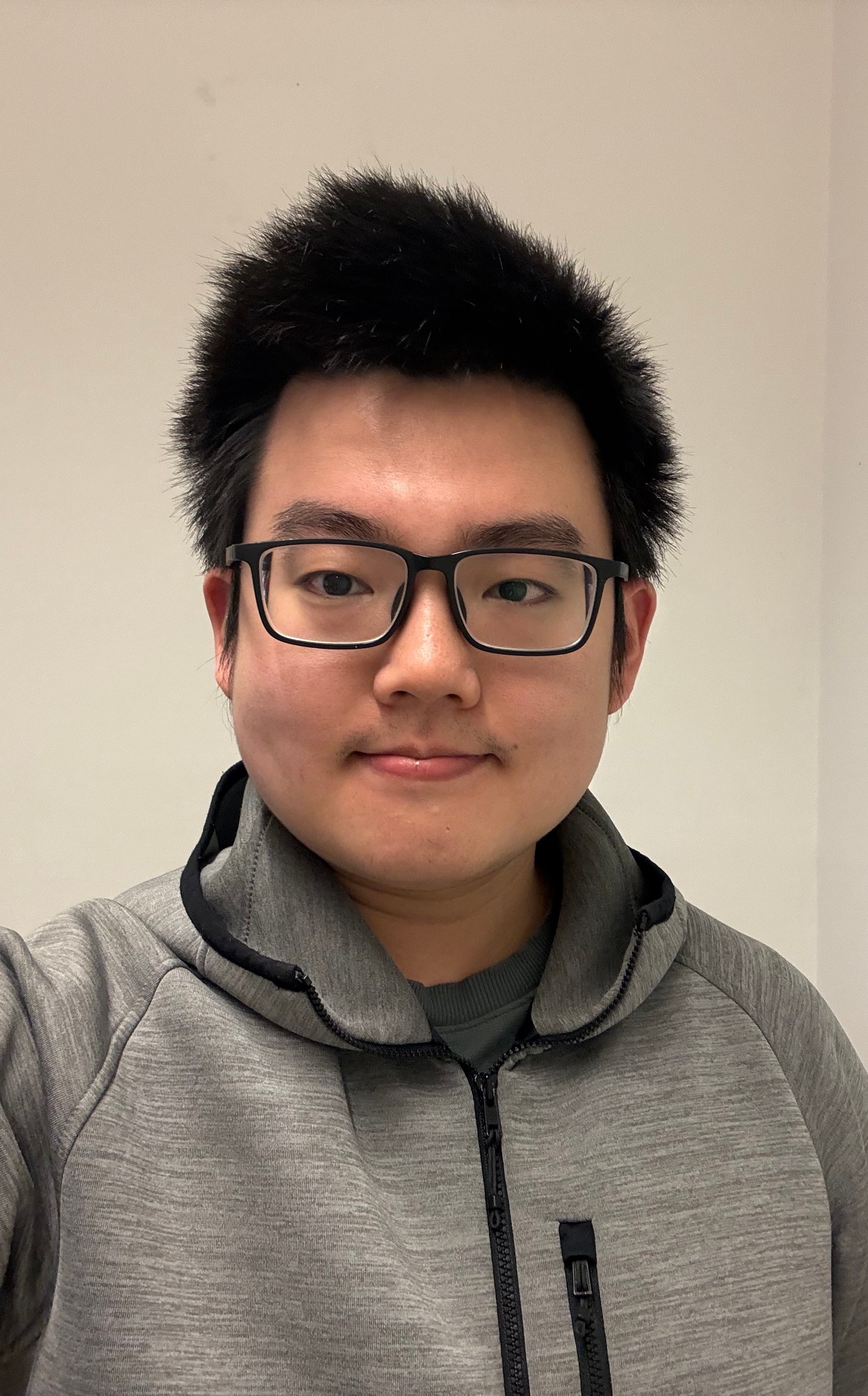}}]{Daniel Wang} is a Ph.D. student in the Distributed Systems Group at TU Wien. His main research interests include deep reinforcement learning, active inference, and self-adaptive distributed systems.
\end{IEEEbiography}

\vspace{-1.5cm}

\begin{IEEEbiography}[{\includegraphics[width=1in,height=1.25in,clip,keepaspectratio]{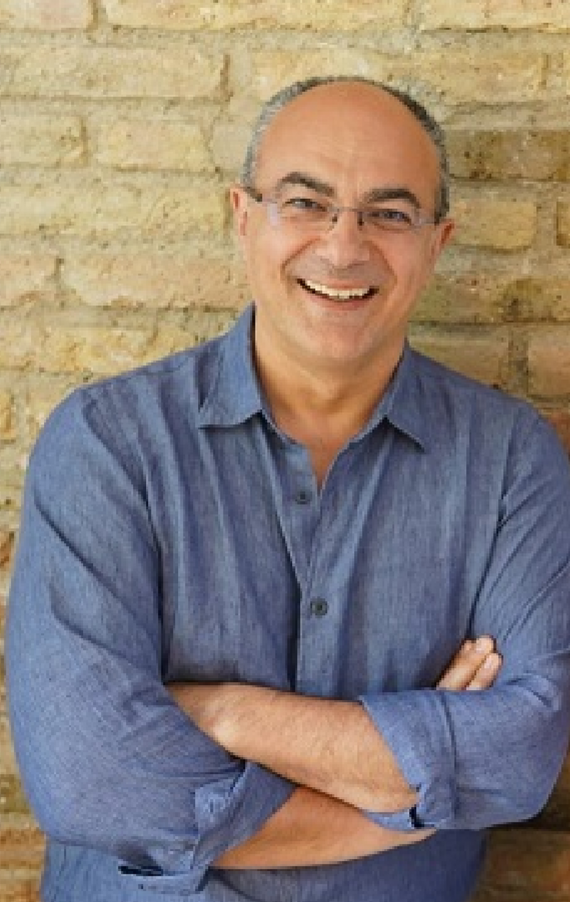}}]{Schahram Dustdar} is a Full Professor of Com-
puter Science (Informatics) with a focus on Internet
Technologies heading the Distributed Systems Group
at the TU Wien. He was founding co-Editor-in-
Chief of ACM Transactions on Internet of Things
(ACM TIoT). He is Editor-in-Chief of Computing
(Springer). He is an Associate Editor of IEEE Trans-
actions on Services Computing, IEEE Transactions on
Cloud Computing, ACM Computing Surveys, ACM
Transactions on the Web, and ACM Transactions on
Internet Technology, as well as on the editorial board
of IEEE Internet Computing and IEEE Computer. Dustdar is recipient of
multiple awards: TCI Distinguished Service Award (2021), IEEE TCSVC
Outstanding Leadership Award (2018), IEEE TCSC Award for Excellence
in Scalable Computing (2019), ACM Distinguished Scientist (2009), ACM
Distinguished Speaker (2021), IBM Faculty Award (2012). He is an elected
member of the Academia Europaea: The Academy of Europe, where the
chairman of the Informatics Section for multiple years. He is an IEEE Fellow
(2016), an Asia-Pacific Artificial Intelligence Association (AAIA) President
(2021) and Fellow (2021). He is an EAI Fellow (2021) and an I2CICC Fellow
(2021). He is a Member of the IEEE Computer Society Fellow Evaluating
Committee (2022 and 2023).
\end{IEEEbiography}

\end{document}